\documentclass[pre,twocolumn,floatfix]{revtex4}%

\usepackage{graphicx}
\usepackage{dcolumn}
\usepackage{bm}

\usepackage{algorithmic}
\usepackage{algorithm}
\usepackage{subfigure}
\begin{document}

\preprint{APS/123-QED}

\title{Let Your {\em Cyber}Alter Ego Share Information and Manage Spam}

\author{Joseph S. Kong}
\email{jskong@ee.ucla.edu}
 \affiliation{Electrical Engineering Department \\ University of
California, Los Angeles.}
\author{P. Oscar Boykin}%
\email{boykin@ece.ufl.edu}
\affiliation{Electrical and Computer Engineering Department \\
University of Florida.}

\author{Behnam A. Rezaei}
\email{behnam@ee.ucla.edu}
\affiliation{Electrical Engineering Department \\ University of
California, Los Angeles.}
\author{Nima Sarshar}
\email{nima@ee.ucla.edu}
\affiliation{Electrical Engineering Department \\ University of
California, Los Angeles.}
\author{Vwani P. Roychowdhury}
\email{vwani@ee.ucla.edu}
\affiliation{Electrical Engineering Department \\ University of
California, Los Angeles.}
\date{\today}

\begin{abstract}
Almost all of us have multiple cyberspace identities, and these
{\em cyber}alter egos are networked together to form a vast
cyberspace social network. This network is distinct from the
world-wide-web (WWW), which is being queried and mined to the tune
of billions of dollars everyday, and until recently, has gone
largely unexplored. Empirically, the cyberspace social networks
have been found to possess many of the same complex features that
characterize its real counterparts, including scale-free degree
distributions, low diameter, and extensive connectivity. We show
that these topological features make the latent networks
particularly suitable for explorations and management via
local-only messaging protocols.  {\em Cyber}alter egos can
communicate via their direct links  (i.e., using only their own
address books) and set up a highly decentralized and scalable
message passing network that can allow large-scale sharing of
information and data.  As one particular example of such
collaborative systems, we provide a design of a spam filtering
system, and our large-scale simulations show that the system
achieves a spam detection rate close to 100\%, while the false
positive rate is kept around zero. This system of letting {\em
cyber}alter egos network among themselves has several advantages
over other recent proposals for collaborative spam filtering:  (i)
It {\em uses an already existing network}, created by the same
social dynamics that govern our daily lives, and no dedicated
peer-to-peer (P2P) systems or centralized server-based systems
need be constructed; (ii)  It utilizes a {\em percolation search}
algorithm (which can be viewed as mimicking how rumor is spread in
a social network) that makes the query-generated traffic scalable;
(iii) The network has a {\em built in trust system} (just as in
social networks) that can be used to thwart  malicious attacks;
and (iv) It {\em can be implemented right now} as a plugin to
popular email programs, such as MS Outlook, Eudora, and Sendmail.

 \end{abstract}

\pacs{Valid PACS appear here}
\maketitle
\section{ Introduction}\label{sec:intro}

\subsection{ {\em Cyber}Alter Ego and the Pervasive Cyberspace Social Networks}
Our socioeconomic activities are  getting intricately entwined
with our identity in the cyberspace, and perhaps we are witnessing
the emergence of an alter ego in the cyberspace. For example,
every email user can construct a list of email addresses from
which he has received emails or sent emails to; this constitutes
one's {\em cyber-neighborhood}.  This list is stored in the
address books or contact lists managed by one's email client
software or by the ISPs that one uses. It can be also
automatically constructed by just sifting through one's mail box.
Individuals on such lists have their own address books and contact
links and soon there is a cyberspace network, in which our
identities or {\em cyber}alter egos are firmly embedded and occupy
various positions of power,  centrality, or proximity to
cyber-communities of potential interest. Thus, an {\em undirected
social email network} can be defined as follows: the nodes in the
network correspond to email addresses; a pair of nodes is
connected by an edge if a message is exchanged between the two
nodes. Similarly, a {\em directed social email network} can be
defined as follows: nodes also correspond to email addresses; a
directed edge points
 from $A$ to $B$ if node $A$ has sent an email to node $B$ and vice versa\footnote{In this paper,
we use the term social email network in the sense of an undirected
network, except when we discuss the computation of trust.}. One
can  modify this network to incorporate other parameters of
interest; for example,   each edge can be assigned a weight based
on the number of email messages exchanged, or time-stamps can be
added to messages along each edge so that one can prune the
network to reflect the recent status of interactions among the
{\em cyber}alter egos.

A major obstacle to studying such email networks has been that
contact addresses and lists of a large enough group of {\em
cyber}alter egos are not available in the public domain.  Even
though large ISPs, such as Hotmail, Yahoo, and AOL, have this
information for all their users, they are not for public
consumption. Drawn by the commercial potential of these latent
networks, a number of companies \cite{spokes} have started
providing services where participants can upload their address
books, allowing the corporation to create a central server where
the social email network is stored and updated; the goal is to
provide services to the participating clients by mining the
network. These networks, however, are also proprietary, because of
both privacy and commercial secrecy reasons. Fortunately for us
(i) the system that {\em we have designed do not make use of the
knowledge of the complete network}; to carry out the protocols
described here, the {\em cyber}alter egos have only to exchange
messages with those on their  own contact lists and do not have to
know about its cyber-neighbor lists, and (ii) A few  examples of
social email networks have been thoroughly studied in the
literature, allowing us to observe that {\em they share many of
the same complex features as} real world social networks. In
particular, we will use the network  analyzed by Ebel et. al. in a
recent work\cite{Ebel}, which shows that the network has a
scale-free structure, short diameter, and a giant connected
component (gcc) that contains more than 95\% of the nodes.

Since our {\em cyber}alter egos are becoming more entrenched as a
significant part of our overall social and commercial selves,
can one start managing and utilizing their network the same way
that we manage our real-life social networks? Any such effort
should abide by rules, such as the need to protect the privacy of
the users and also the need to allow  participants to dynamically
decide whether they want to participate or not. The primary
contribution of this paper is to {\em provide a decentralized,
efficient, and scalable system for querying and sharing
information on the global social networks}. One {\em major
application} of this overlay information management system is to
{\em filter spam}, as reviewed in the following.

\subsection{ Spam and Content-based Spam Filtering}
Spam, or Unsolicited Bulk Email, is plaguing internet users around
the world.  It has been estimated that approximately 68\% of the
worldwide email traffic today is spam and up to 87\% of the emails
directed to US users is spam\cite{SpamStat}.

For the past few years, numerous spam filters have been proposed
and deployed, and of all the existing anti-spam solutions, two
classes of spam filters have emerged as the most effective and
widely-deployed: \emph{Bayesian/rule-based} spam filters and
\emph{collaborative} spam filters. A Bayesian filter uses the
entire context of an e-mail in looking for words or phrases that
will identify the e-mail as spam based on the experiences gained
from the user's sets of legitimate emails and
spams\cite{Rigoutsos04}. One example of a widely deployed Bayesian
spam filter is SpamAssassin\cite{Spamassassin}. Although the
Bayesian anti-spam solutions offer very impressive performances,
they suffer from several serious drawbacks: first, Bayesian
filters require an initial training period and exhibit a downgrade
in performance for responding to messages composed of previously
unknown words; second, Bayesian filters are unable to block
messages that do not look like a typical spam such as messages
that is consist of only a URL or messages that are padded with
random words. Most recently a number of multifaceted approaches
have been proposed\cite{Segal04,Leiba04}. They consider combining
various forms of filtering with infrastructure changes, financial
changes, legal recourse, and more, to address shortcomings of
regular statistical filters.

\subsection{ Collaborative Spam Filtering: Prior Work and Challenges}\label{sec:related}
The increasing realization that the dynamic of spam constitutes a
complex phenomenon brewed, fostered and propagated in the
interconnected realm of the cyberspace, has prompted the use of
collaborative spam filters, where the basic idea is to use the
collective memory of, and feedback from, the users to reliably
identify spams.  That is, {\em for every new spam that is sent
out, some user  must be the first one to identify it} upon
receiving this spam (e.g., by using a Bayesian filter or locally
generated white and black lists); now, any subsequent user that
receives an email that is a suspect,  can query the community of
email users to find out if it has been already tagged as spam or
not. In contrast to Bayesian type filters, collaborative spam
filters do not suffer from the drawbacks just mentioned above, and
it has been shown that they are also capable of superior spam
detection performance\footnote{SpamNet\cite{Spamnet}, a fairly
popular commercial collaborative anti-spam system, claims that a
detection rate closed to 100 \% and a false positive rate closed
to 0 \% are achieved.}.
 {\bf The existing collaborative filtering schemes mostly ignore the already present and pervasive social communities in the cyberspace} and try to create new communities of their own to facilitate the sharing of information.  This unenviable task of creating new social communities is beset with several difficulties that have limited the deployment and effective use of most collaborative filtering schemes proposed so far. The challenges include:\\
{\bf (i)} \emph{How to find users to participate?:} In order for a
collaborative spam filter to be highly effective, a large number
of users (on the order of hundreds of thousands or millions) must
be participating in using the system. However, effectively finding
and interconnecting a large number of willing participants is
non-trivial. In other words making any artificially established community acceptable and popular is an unpredictable and difficult task at best, and impossible at worst. \\
{\bf (ii)} \emph{How to make the search scalable?}: The power of a
collaborative spam filter lies in the fact that spam data
resources from a large number of users are pooled together and
utilized to fight spam.  In order to avoid high server cost, the
spam databases are typically stored locally on users' computer.
Finding a way to do efficient searches on a network of distributed
databases is very
challenging.\\
{\bf (iii)} \emph{Who to trust}: Inevitably, there would be
malicious users who try to subvert the collaborative anti-spam
system by providing false information regarding spam.  Therefore,
a trust scheme must be devised to place more weights on the
opinions of some provably trustworthy users than on some unknown
users who can be potentially malicious.

The different proposed schemes for collaborative filtering attempt
to address the above challenges to different degrees of
effectiveness. For example, SpamNet\cite{Spamnet} employs the
following mechanisms to address the challenges stated above: It
{\em uses a central server model} to connect all the willing
participants of this collaborative spam filter.  {\em The central
server solution is not scalable} as the system scales and the
server becomes a single point of attack or failure. In addition,
SpamNet employs a complicated algorithm to compute the trust score
for each of its user.  SpamWatch\cite{Zhou} is a totally
distributed spam filter based on the Distributed Hash Table (DHT)
system Tapestry\cite{Zhao}. SpamWatch addresses the three
challenges of a collaborative spam filter in the following ways:
First, SpamWatch uses a DHT-based P2P system to connect all the
participants.   The primary drawback in using a DHT for
collaborative spam filtering purpose is that DHT's do not provide
a natural platform to network existing databases, such as every
email user's personal database of spams. Merging and mining
existing sets of databases is very difficult if not impossible.
Second, SpamWatch uses a hash-based mechanism called Approximate
Text Addressing (ATA) to perform general query searches for spams.
However, as seen in the description of the ATA algorithm, {\em
supporting general query search in a DHT is very complex and
involves expensive operations.} DHT's typically excel at exact-match
lookups but does not perform well for application that needs to
support general search queries such as in a collaborative spam
filter.  Lastly, SpamWatch does not offer any mechanism to address
the trust issue. Most recently Gray \emph{et. al.} have proposed
CASSANDRA, a collaborative spam filter where the network is formed
as clusters of trusted and similar peers. Finally a new reputation
analysis have been proposed by Golbeck \emph{et. al.}
\cite{Golbeck04} where reputation relationships are inferred from
the structure and are used as a method to score emails.

\subsection{ Harnessing The Global Social Email Network}
Recently, Boykin and Roychowdhury investigated the notion of
utilizing social network to do spam filtering\cite{Boykin}. In
their work, it was shown that just by looking at the clustering
coefficient of an email user's personal contact networks, their
algorithm is able to achieve a spam detection rate of 53\% {\em
with zero false positives}. Although this algorithm is very
attractive, it ignores the larger social email network and focuses
only on a projection of it, as witnessed by an individual user,
and it begs the questions whether the larger social email networks
can be harnessed.

In this paper,  we show that a high-performance, scalable and
secure information management and query system can be overlaid on
the social email networks, and provide a case study for
collaborative spam filtering.  The basic idea is the same as that
of other proposed collaborative spam filters; however, instead of
using specialized network, we use the latent social email network
over which the queries and messages are exchanged. We show how
{\em the three challenges outlined in the preceding discussions}
can be {\em effectively addressed} using the topological
properties of the underlying social email networks and {\em recent
advances in complex networks theory}.  {\bf First, } no especially
designed network has to be created for collaborative filtering. In
fact, one of the main features of this system is that {\em all
queries and communications are exchanged via email} through
personal contacts, and that {\em no server or a traditional P2P
system with TCP/IP connections is needed.} {\bf Second,} we
observe that social email networks correspond to  Power-Law (PL)
graphs\footnote{For a power-law (PL) degree distribution, the
probability of a randomly chosen node to have degree $k$ scales as
$P(k)\propto k^{-\gamma}$ for large $k$; $\gamma$ is referred to
as the exponent of the distribution.} \cite{Ebel}, with a PL
coefficient around $2$.  Hence, {\em the underlying network
naturally possesses a scale-free structure} that is a key
hall-mark of many unstructured P2P systems that have organically
grown for file-sharing on the Internet. One can  then utilize a
scalable global search system, namely the {\em percolation search
algorithm}, recently proposed by Sarshar et. al. \cite{Sarshar},
on this naturally scale-free graph of social contacts to enable
peers to exchange their spam signature data. {\bf Third,} one can
{\em harvest and utilize the trust that is embedded in the web of
email contacts}.  By regarding contact links as local measures of
trust and using a distributed Singular-Value-Decomposition (SVD)
algorithm, we can obtain a trust score called \emph{mailtrust}. In
fact, the famous Google PageRank\cite{Brin} is computed in a
similar fashion. {\bf Finally,} the proposed system {\em can be
implemented right now} as plugin to popular email programs, such
as the MS Outlook.

We show via extensive simulations that the system is also {\em
capable of delivering high performances while incurring minimal
costs.}  Under the assumption that there would be a large number
of users (on the order of hundreds of thousands or millions), the
system can offer a spam detection rate around 99\%; in fact, the
detection rate can reach close to 100\% when the number of users
approach the internet scale. At the same time, the number of false
positives in our system can be tightly controlled to a level very
close to zero. Meanwhile, {\em as the number of users of the
system scales, the communication cost of the system would be kept
at a sublinear scale and the memory storage cost would grow only
at a logarithmic scale}. In addition,
 due to the fact that {\em no TCP/IP connection is required} and all
communications in the system is done via background email
exchanges, {\em less computational and networking burden would be
placed on local computers.}  Lastly, the system is designed to be
secure and rigorously protective of users' privacy and
confidentiality.

The rest of the paper is organized as follows.  In section
\ref{sec:background}, we present the background theory and the
important concepts vital to this paper, such as email network
theory and the percolation search algorithm.  In section
\ref{sec:architecture}, we describe the protocol of our social
network based collaborative anti-spam system in detail.  In
section \ref{sec:simulation}, we use a real world email network to
perform large-scale simulations of the system. In section \ref{sec:threat},
we construct a threat model and show by simulation that a social
network based trust scheme is effective in minimizing damages
caused by malicious users.  Finally, in section
\ref{sec:misc}, we address several important topics such as the
protection of privacy and the system's resilience against random
user failure.

\section{Background Concepts}\label{sec:background}
Our system is motivated by a number of recent advances in {\em
complex networks theory and systems}, Eigen-methods based
computation of trust and relevance, and the proven efficacy of the
spam digest system as signatures of emails.  We briefly review
this background material in this section.

\subsection{ Topology of Social Email Networks}\label{ssec:social_network}
A particular email network comprising 56,969 nodes (i.e., email
addresses) has been studied by Ebel et. al.\cite{Ebel}  Based on
the statistics reported in Ebel's work, we identify three
desirable properties that would make social email networks an
attractive platform for building a collaborative spam filter:\\
{\bf (i)} An email network has been found to possess a scale-free
topology.  More precisely, for the email network examined in
\cite{Ebel}, the node degree distribution follows a power law
(PL): $ \displaystyle P(k) \propto k^{-1.81},$ where $k$ is the
node degree, and $P(k)$ denotes the probability
that a randomly chosen node has degree equal to $k$. One of the consequences of this property is that of very low percolation threhold\cite{Sarshar}; in other words, {\em the network is extremely resilient to random deletions of nodes}.  One can also show that even if high-degree nodes are deleted preferentially, {\em one has to remove almost all the high-degree nodes}, before the network gets fragmented. \\
 {\bf (ii)} A large
fraction of the nodes ($\tilde 95.2\%$) in a social email network
is connected to the giant connected component (GCC).  This means
that any node can reach almost any other arbitrary node by simply
following email links.\\
{\bf (iii)} The email network has a low diameter (i.e. there exist
short paths between almost any pair of two nodes in the network).
In fact, for the email network investigated by Ebel et.
al.\cite{Ebel}, the mean shortest path length in the giant
connected component was found to be $l=4.95$ for a component size
of $56,969$ nodes.  This short-diameter property allows any email
user to efficiently communicate with any other email user in the
network by crossing only a few email contact links.

The above properties of the social email network should not come
as a surprise, since it reflects the same social dynamics that we
practice in our everyday life.
\begin{figure}[tbh]
 \subfigure[][\label{fig:percolation_plot}]{
 \centering
\includegraphics[width=3.1in,height=2.4in]{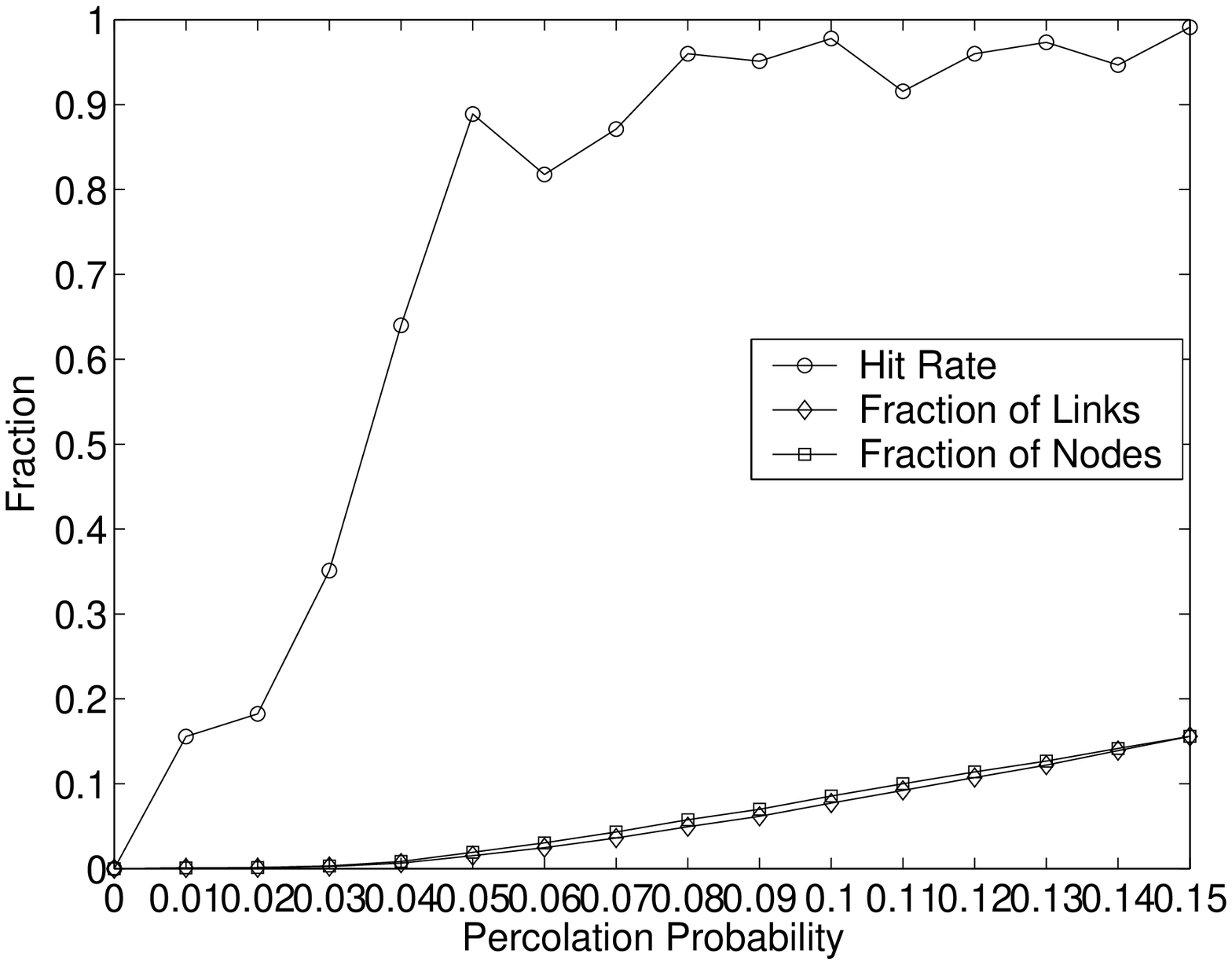}
}
 \subfigure[][]{ \centering
\includegraphics[width=3.1in,height=2.4in]{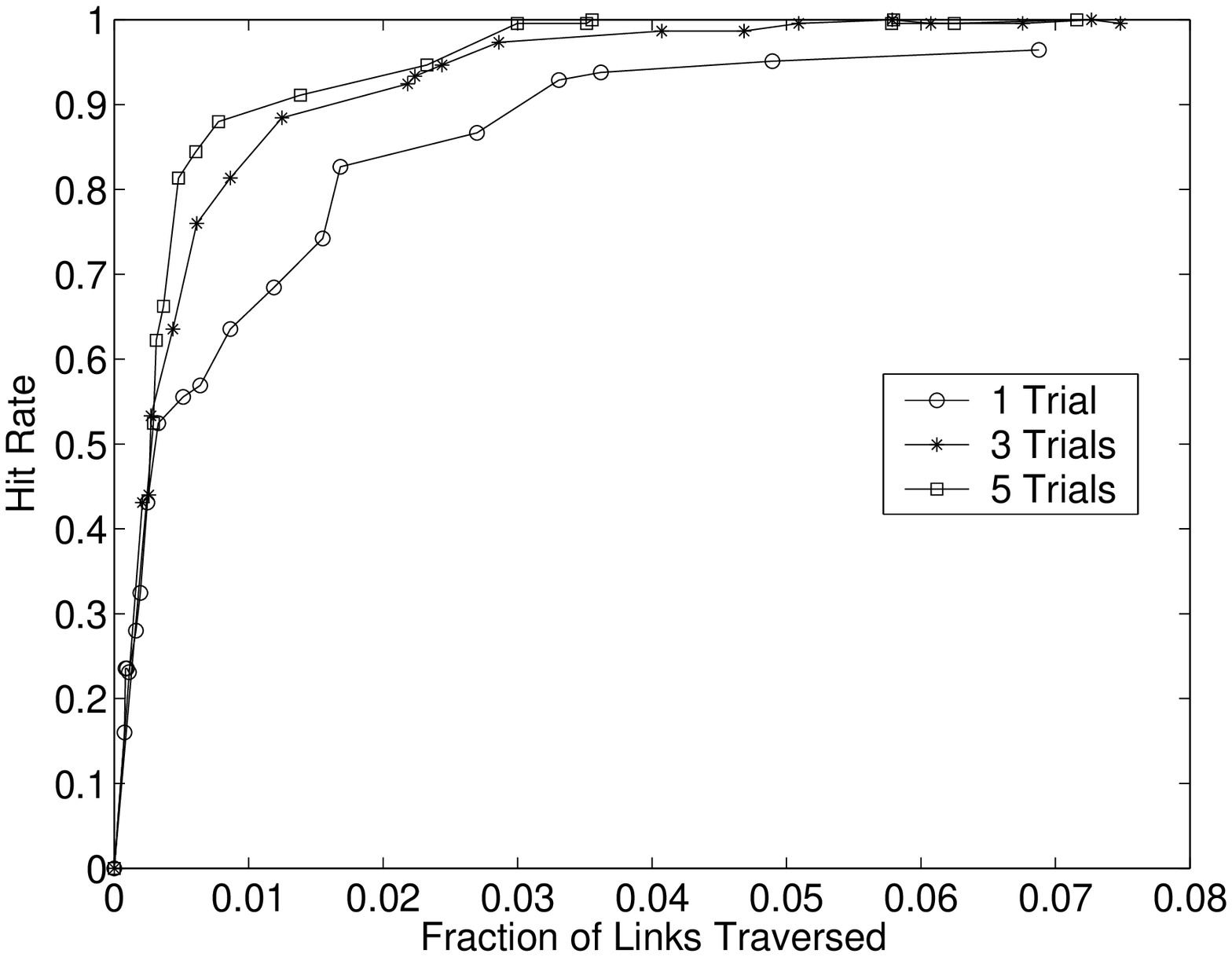}
} \caption{\small \textbf{Percolation Search On Social Email
Networks:} (a){\em The hit rate}, fraction of links and fraction
of nodes traversed as a function of the percolation probability.
Notice that there is a sudden jump in the hit rate above the
percolation threshold, while the fraction of links and nodes
processing the search query increases only linearly, after the
threshold. The network used in this percolation search simulation
is a real-world email contact network. The number of nodes is
56,969, $\tau \approx 1.81$, the TTL is 50 for both query and
content implants and only one unique content exists in the
network. (b) Hit rate for percolation search on email contact
network with TTL of $50$. {\em Repeating the percolation trial
multiple times pushes the hit rate exponentially closed to 1.}}
\label{fig:percolation_trials}
\end{figure}

\subsection{ Percolation Search and Scalability}\label{ssec:percolation}
We can utilize the percolation search algorithm proposed by
Sarshar et. al.\cite{Sarshar} that exploits the presence of a
tightly connected core comprising mostly high-degree nodes. In
particular, {\em it is shown in \cite{Sarshar} that unstructured
searches in PL networks can be made highly scalable} using the
percolation search algorithm. The algorithm involves message
passing on direct links only, and in some sense it resembles how
rumors propagate in social networks. The key steps of the
algorithm are as follows:\\
{\bf (i)} \emph{Caching or Content Implantation}: Each node
performs a short random walk in the network and caches its content
list on each of the visited nodes.  The length of this short
random walk
is specified later and is referred to as the Time To Live (TTL). \\
{\bf (ii)} \emph{Query Implantation}: When a node intends to make
a query, it first executes a short random walk of the same length
as step 1 and implants its query requests on the nodes visited.
The length of this random walk is usually taken to be the same as
the
TTL used in the content implantation process. \\
{\bf (iii)} \emph{Bond Percolation}: All the implanted query
requests are propagated through the network in a probabilistic
manner; upon receiving the query, a node would relay to each of
its neighboring nodes with percolation probability $p$, which is a
constant multiple of  the percolation threshold, $p_c$, of the
underlying  network.

It is shown in \cite{Sarshar} that the percolation threshold of
any random network is given as $p_c=\langle k\rangle/\langle
k^2\rangle$. For a PL network with exponent $\tau$ and maximum
degree $k_{max}$, we have  $\langle k^2\rangle =
O(k_{max}^{3-\tau})$ and  $\langle k\rangle =
O(k_{max}^{2-\tau})$, and hence, we get a percolation threshold of
$p_c=O(k_{max}^{-1})$, which is vanishingly small if $k_{max}$
increases with the size of the network, which is usually the case.
Thus, if we percolate at a multiple $\gamma$ of $p_c$, then the
total traffic generated would be, $\mathcal{C_\tau}=\gamma p_c
\langle k\rangle N= O(\frac{\langle k\rangle^2 N}{\langle
k^2\rangle})= O\left(k_{max}^{-\tau + 1}N\right)$. In real world
networks, $k_{max}$ typically scales sublinearly as a function of
the network size. For $k_{max}=O(N^{1/\tau})$, we have:
$\mathcal{C}_\tau=O\left(k_{max}^{-\tau +
1}N\right)=O(N^\frac{1}{\tau})$. For a detailed analysis of the
hit rate and how it behaves as one performs multiple searches see
\cite{Sarshar}.

Since the social email networks have a PL degree distribution, it
is ideally suited for reaping the benefits of a percolation
search, and the simulation plots obtained from performing
percolation search on the real email dataset \cite{Ebel, EbelData}
are provided in Fig. \ref{fig:percolation_trials}.

%

\subsection{ The MailTrust Algorithm}\label{ssec:trust}
{\em Just as in the case of WWW, where the PageRank captures the
relevance of a particular web page, the topological structure of
the social email networks can be used  to assign trust or
reputation to individual users.} First, we model each email
contact as placing a unit of trust on the recipient.  Thus, for a
node that contacts $k_{out}$ other nodes, we can compute the
fraction of trust that this node places on each of his
out-neighbors as followed: the trust for neighbor $i$, $t_i$, is
equal to the number of emails sent to neighbor i divided by the
total number of emails sent.  Note that the collection of $t_i$'s
forms a probability vector, called the personal trust vector
$\overrightarrow{t}$. Thus, if we model the entire email network
as a discrete time Markov chain, the local trust vector,
$\overrightarrow{t}$, becomes the transition probability function
for each node. We then compute the steady state probability vector
using  Power Iteration method which is the the same algorithm
adopted to compute pagerank score of documents on web
\cite{Kamvar,Brin}. As discussed in the literature, one needs to
make sure that this Markov chain is ergodic and this can be
achieved by having nodes with zero out-degree assign uniform trust
to a set of pre-trusted nodes who have been carefully picked. An
alternate way to compute the MailTrust score in a distributed
fashion can be found in \cite{Kamvar}, along with a scheme on how
the trust scores can be kept securely in the system even with the
presence of malicious users.

\begin{figure}[htb]
\centering
\includegraphics[width=3.3in,height=2.5in]{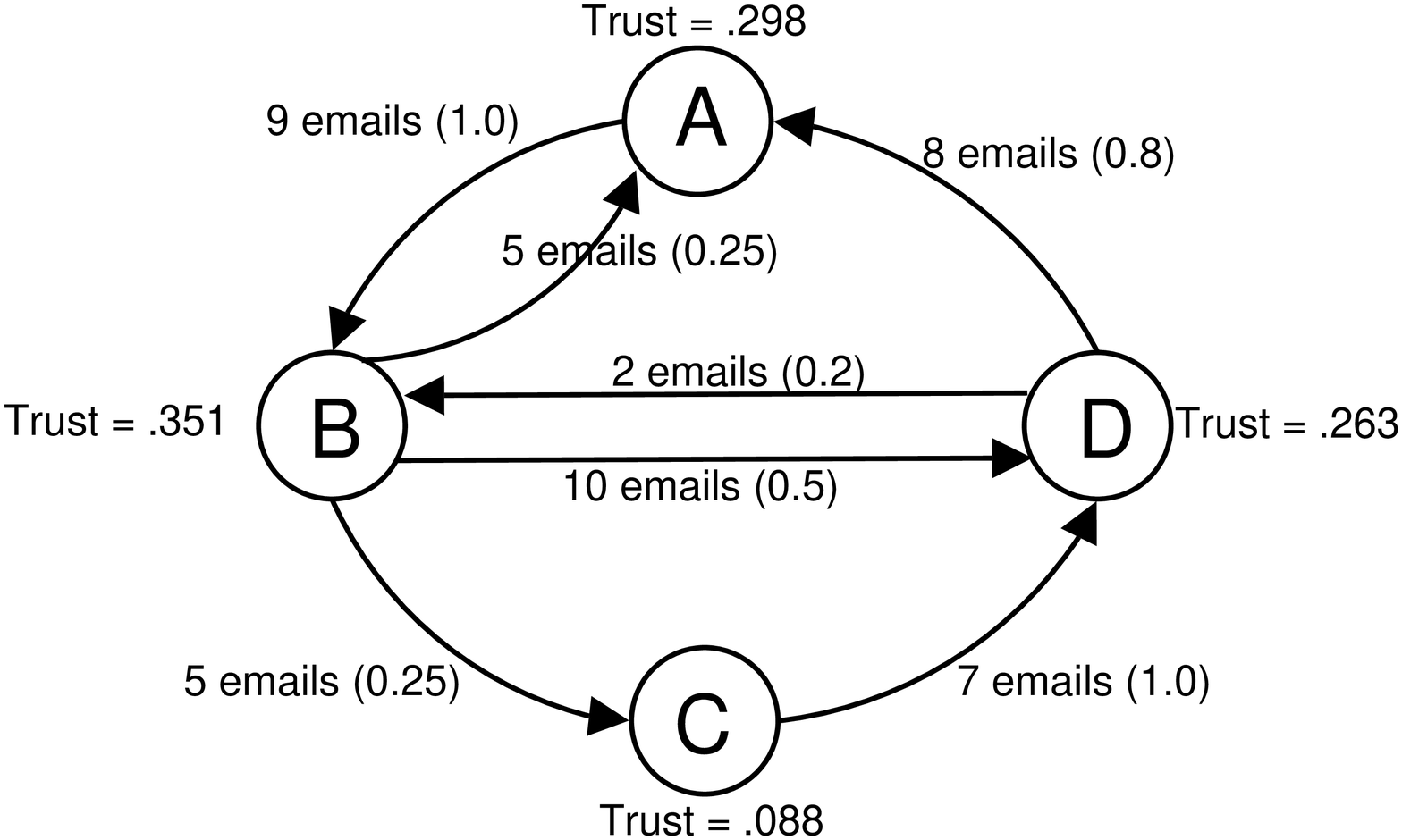}
\caption{\small \textbf{MailTrust:} A simple illustration of the
MailTrust algorithm. The numbers in parentheses represent the
local trust values that each node places on his/her neighbors. The
MailTrust scores for each node is then obtained by computing
steady state probability vector of the Markov chain. }
\label{fig:mail_trust}
\end{figure}

\begin{figure}[htb]
\centering
\includegraphics[width=3.3in,height=2.5in]{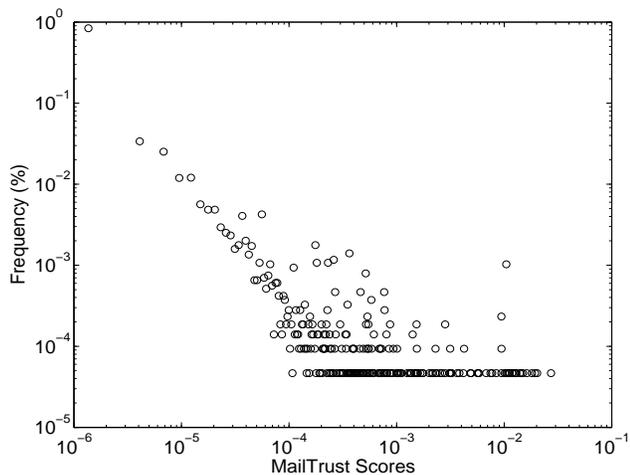}
\caption{\small\textbf{MailTrust Distribution:} The probability
density function of MailTrust scores using $10^4$ bins. These
scores are obtained by applying the MailTrust algorithm on the
email network data set from \cite{Ebel}. Notice that this
probability density function is heavy-tailed, indicating that a
few nodes are much more trustworthy than most nodes.}
\label{fig:trust_plot}
\end{figure}

We will refer to this trust score as \emph{MailTrust} in the rest
of this paper. A plot of the MailTrust scores obtained from
\cite{Ebel, EbelData} is shown in Fig. \ref{fig:trust_plot}.

\subsection{ Digest-based Spam Indexing}\label{ssec:digest}
In a collaborative spam filtering system, it is important to have
an effective mechanism to index known spams so that subsequent
arrivals of the same spam can be correctly identified.  The
collaborative design of the system does not depend on any specific
algorithm, but for initial experimental results we have adopted
the well known \emph{digest-based indexing} mechanism
\cite{DamianiDigest} to share spam information between users.
  Damiani et. al. have rigorously demonstrated that
the digest algorithm described in \cite{DamianiDigest} is highly
resilient against the possible forms of automatic modifications of
spam emails. The digest algorithm is further shown to satisfy both
the privacy preserving and that it produces almost close to zero
false positives (i.e., the digest of one email matches the digest
of an unrelated email).

\section{Implementation and System Protocol}\label{sec:architecture}
In order to use our proposed collaborative spam filtering system,
an interested individual must first obtain a simple client program
that works as a plug-in to an email program such as MS Outlook,
Eudora, Sendmail, etc\footnote{However, implementing the client
program as an email program plug-in is not the only option; large
email providers can also implement this system on the email server
ends.}.  This simple client will only need to provide the
following features: first, the client must come with a
digest-generating function as specified in section
\ref{ssec:digest}; second, the client is responsible for keeping a
personal blacklist of spams for the end-user as well as caching
blacklists of spams for other nodes as described in the section on
the percolation search algorithm, (see section
\ref{ssec:percolation}); third, the client would have access to
the list of social email contacts (both inbound and outbound) of
 the end-user. The pseudo code of the distributed client is given in Algorithm
 \ref{Alg:ProcessMail}.

  \begin{algorithm}
\caption{PROCESS-MAIL(Email $E$)} \label{Alg:ProcessMail}
\begin{algorithmic} [1]
 \IF{DefinitelySpam($E$)}
    \STATE Mark $E$ as Spam
 \ELSIF{DefinitelyNotSpam($E$)}
    \STATE Mark $E$ as not Spam
 \ELSE

    \STATE $D_e$ = Digest(E);\COMMENT{Gray SPAM};
    \STATE Implant percolation of $D_e$ on a random walk of length
    $l$
    \STATE Wait(T);
    \STATE $H_e = $  HitScore();
    \IF { $H_e<threshold$}
    \STATE Mark $E$ as not Spam
    \ELSE
    \STATE Mark $E$ as spam
    \ENDIF
 \ENDIF
\end{algorithmic}

\end{algorithm}

\begin{algorithm}
\caption{Publish-Spam(Email $E$)} \label{Alg:Publish}
\begin{algorithmic}[1]
\STATE $D_e$ = Digest(E); \STATE Implant $D_e$ on a random walk of
length $l$
\end{algorithmic}

\end{algorithm}
\begin{algorithm}
\caption{HitScore(Hits)} \label{Alg:Hitscore}
\begin{algorithmic}[1]
\IF {Using MailTrust}
    \STATE HitScore $=\Sigma_{h\in\textrm{Hits}}{mailtrust(h)}$
\ELSE
    \STATE HitScore $=|Hits|$
\ENDIF \STATE Return HitScore;
\end{algorithmic}

\end{algorithm}

\begin{figure*}[htb]
\centering
\includegraphics[width=4in,height=6in,clip=false,angle=270]{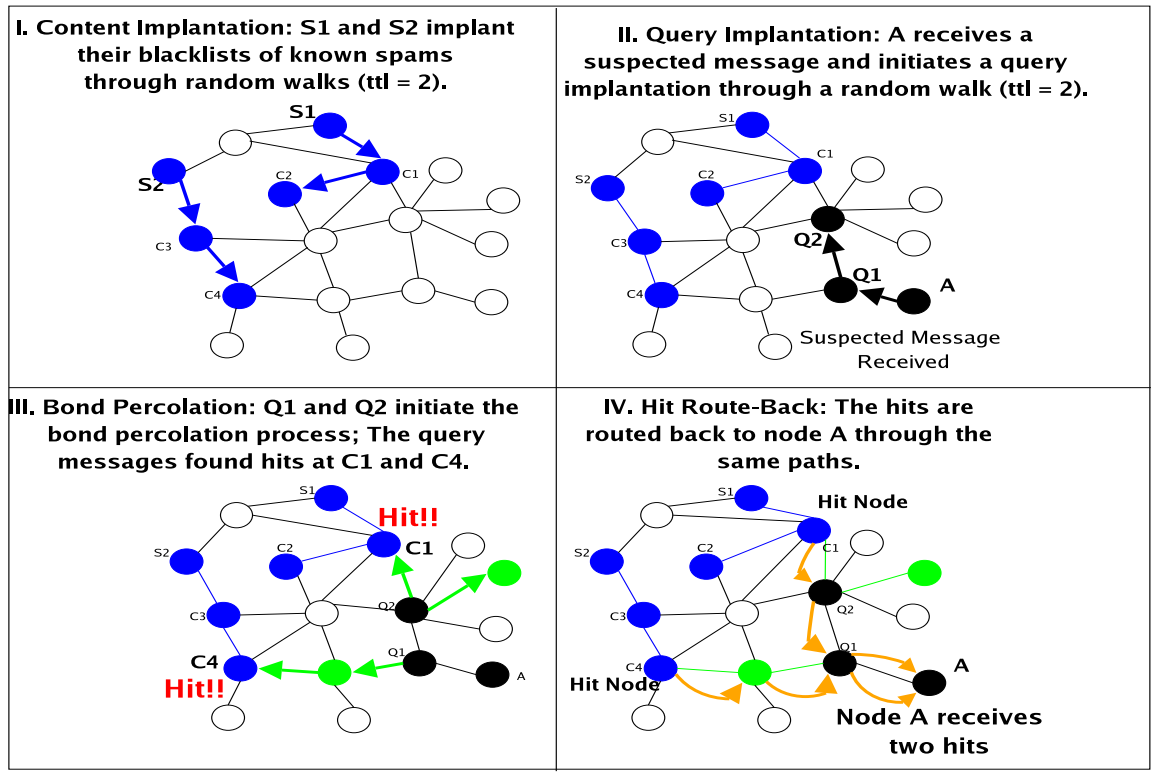}
\caption{An illustration of the protocol of the system.}
\label{fig:protocol}
\end{figure*}

\textbf{Message Arrivals and Digest Indexing:} When an email
message arrives at the end-user, the method checks whether it is
definitely spam or not spam(DefinitelySpam/DefinitelyNotSpam). Any
traditional spam filtering method like white-list, blacklist,
Bayesian filter, etc.  can be integrated to create a hybrid multi
tier architecture. \emph{DefinitelyNotSpam} for example can be a
white list of addresses in the contact list and
\emph{DefinitelySpam} can be output of a Bayesian filter when the
filter indicates email as spam with very high probability. If an
email is then suspected to be spam, the client program would call
the digest function to generate a digest, $D_e$, for the message.

\textbf{Making a Query in the System:} Now, we would query the
system to find out whether any other user in the network already
has the digest, $D_e$, on its spam list. Each query message for
this digest is then implanted on a random walk of length $l$.
Nodes with an implanted query request will then percolate the
query message containing the digest, $D_e$, through their email
contact network using a probabilistic broadcast scheme as
specified in the \emph{bond percolation} step of the percolation
search algorithm. Each node visited by the query would declare a
hit if the digest, $D_e$, matches with any of the digest that is
cached on that node\footnote{Please refer to Damiani et. al.'s
work \cite{DamianiDigest} on the definition of matching digests.}.
All the hits would be routed back to the node that originated the
query through the same path that the query message arrives at the
hit node. If the nodes have trust scores, then returned hits
include the their trust score as well.


\textbf{Processing the Hits and Making the Decision:} After all
the hits are routed back, \emph{HitScore} is then calculated as
(or as the weighted sum if using trust scheme; see
Section~\ref{ssec:trust}) sum of all the positive hits. If it
exceeds a constant threshold value, the message in question is
declared as spam; otherwise, the email message is determined to be
non-spam.

\textbf{Publishing Digest:} If an email is declared as spam, and
placed in the user's "spam" folder then the \emph{Publish-Spam}
function would be called that generates the digest of the spam
message, $D_e$ and caches the digest on a short random walk, as
specified in the \emph{caching or content implantation} step of
the percolation search algorithm.

\textbf{System Maintenance:} If the EigenTrust algorithm from
section \ref{ssec:trust} is implemented, we would need to update
the trust scores of the nodes on a periodic basis.  Since most
people's amount of email contacts change not faster than a daily
basis, The distributed EigenTrust computation should be performed
at most once a day to obtain up to date trust scores for all
nodes. Connectivity of the network is maintained by simple
background message declaring join/leave sent to each of the user's
contacts.

\section{Simulation and System Performance}\label{sec:simulation}


\textbf{Network Model:} In this section, all simulations are
performed on a real-world email network investigated in Ebel
et.al.'s work\cite{Ebel}. (The email network data can be obtained
via this url \cite{EbelData}.) In the following simulations, only
the \emph{giant connected component} is used, which contains
95.2\% of all nodes in the original dataset. Please see table
\ref{tab:sim_parameters} for the specific values of this email
network's parameters.

\textbf{Spam Arrival Model:} We model the spam detection
performance of a collaborative spam filter as a function of the
number of copies of the similar spam messages that arrive to the
system. In the extreme case that every spam arrived to the system
is unique, one can easily see that a collaborative filter would be
totally futile, since no user can benefit from the prior
identifications of others.

Assuming that similar spam messages are sent to approximately 5
million internet users on average\footnote{No statistics on this
has been found but several online sources suggest that spammers
usually send out on the order of millions of copies per unique
message.} and estimating internet users to be 600 million
worldwide\cite{NumberOnline}; thus, assuming that spammers select
spam targets uniformly randomly from the set of all internet
users, the probability that any individual would receive a copy of
a given spam is approximately 0.8\%.  Since there are 56,969
nodes/users in our email network, the approximate number of
identical spams arrived to this network is about 500.   We further
assume that each spam message arrives at nodes of the network
uniformly randomly.\footnote{We will see that in section
\ref{sec:threat}, this uniform assumption is relaxed. In fact,
when no trust scheme is implemented, the selection of spam target
node has no impact on spam detection performance since all
contents are almost surely to be found with probability
exponentially closed to one.}

\textbf{Specification of Percolation Probabilities:} Recall that
in the percolation search algorithm,  each edge gets a message
with probability $p$ which is chosen to be a constant multiple of
the percolation threshold of the network. In general, the
percolation threshold might not be known, and so one needs to come
up with a scheme to adaptively perform the search using an
increasing sequence of percolation probabilities.  In order to
ensure a high hit rate for queries and a low communication cost
for the system, we propose the following scheme to perform query
searches:  we start the first query with very low percolation
probability; if not enough hits are returned, we send out a second
query with a percolation probability that is twice of the first
one; if still not enough hits are routed back, we repeat the
searches by increasing the percolation probability in this
two-fold fashion until the probability value reaches a maximum
value, $p_{max}$; once this maximum is reached, we repeat the
query with the maximum percolation probability for a constant
number of trials and stop.  The query search is terminated as soon
as the total number of distinct hits routed back reaches the
threshold after any given trial. If {\em no hits are returned
after $n_{rep}$ attempts at the maximum probability $p_{max}$},
then the search is terminated and the queried item is considered
as absent.

For the simulation experiment in this section, we set the starting
percolation probability to be .00625 and $p_{max}$ to be .05, and
$n_{rep}=3$.  All other relevant parameters of the experiment are
specified in Table \ref{tab:sim_parameters}. In addition, we
assume in this simulation experiment that upon the receipt of a
new spam message, all nodes immediately cache this new content on
a random walk as specified in the percolation search algorithm.
This is done regardless of the fact that the spam has been
automatically filtered by the system or it leaked through the
filter and must be identified by human inspection or by some other
means.

\begin{table*}
\centering
\begin{tabular}{||l|l|l||}
\hline Network & \# of nodes & 56,969 \\ \hline
 & \# of edges & 84,190 \\ \hline
& Node degree distribution & Power-Law (PL) \\ \hline & PL
exponent & $\approx$ 1.8 \\ \hline & mean node degree $<k>$ & 2.96
\\ \hline & node degree 2nd moment $<k^{2}>$ & 174.937 \\ \hline &
approximate percolation threshold $(q_{c}) \approx
\frac{<k>}{<k^{2}>}$ & .0169 \\ \hline & time-to-live (ttl) & 50
\\ \hline Simulation Param. & \# of arrivals of the same spam &
500 \\ \hline & threshold (\# of hits needed to identify spam) & 2
\\ \hline & percolation probability trials & [.00625 .0125 .025
.05 .05 $\ldots$] \\ \hline & \# of runs & 30 \\ \hline Threat
Model & \# of time steps & 25 \\ \hline & \# of malicious nodes
inserted per time step & 10 \\ \hline & total \# of mailing lists
& 50,000 \\ \hline & Zipf coefficient & 0.8 \\ \hline & \# of
non-spams queried per time step ($x$) & 1,000 \\ \hline & $m$,
number of items on a blacklist & 10 \\ \hline & \% of user's
non-spam to be queried & 5\% \\ \hline
\end{tabular}
\caption{Simulation Settings} \label{tab:sim_parameters}
\end{table*}


\noindent \textbf{Simulation Execution:} The simulation is
repeated for 30 runs. In each run, 500 copies of the same spam
arrive {\em sequentially} at different nodes in the network.  The
nodes are selected uniformly randomly for each spam message
arrival. The first node receiving it performs a search, but of
course gets 0 hits; similarly, the second node will also get at
most one hit and it will be below the threshold of 2 to be
identified as a spam. For the first two nodes, after the searches
return no hits, the messages are manually tagged as spams. Since
these two initial searches will be considered as misses, {\em the
maximum detection rate} is $498/500=99.6\%$, where the detection
rate is simply the number of successful spam detections divided by
the total number of spam arrivals. We record the the detection
rate for each run, compute the overall average and standard
deviation, and plot the results in error-bar plots. In addition to
the detection rate, we also record the percentage of edges crossed
per query, which is the primary metric for network traffic cost.
We repeat the simulation by varying one parameter: $n_{reps}$,
which is the number of query trials repeated with percolation
probability set at $p_{max}$ before declaring failure.

\begin{figure}[htb]
\centering
\includegraphics[width=3.3in,height=2.5in]{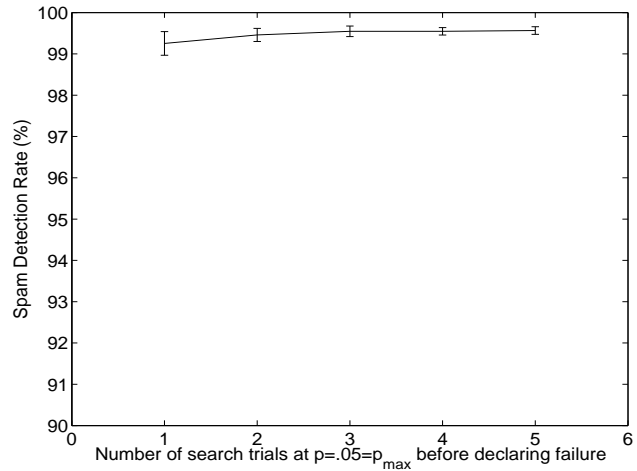}
\caption{\small {\bf Spam Detection Performance:} This figure
plots the simulated spam detection rate (in percentage) as a
function of
 the number of query trials repeated with percolation probability set at $p_{max}$ before declaring
failure.  Note that all the average detection rates are well above
99\%.  The results are averaged over 30 runs and the error bar
plots one standard deviation above and below the mean.}
\label{fig:det_rate_plot}
\end{figure}

\textbf{Simulation Results Analysis:} Fig. \ref{fig:det_rate_plot}
plots the simulated spam detection rate (in percentage) as a
function of $n_{reps}$, averaged over 30 runs.  Note that for
$n_{reps} \geq 3$, the spam detection rate is extremely close to
the maximum detection rate of 99.6\% for this
experiment

\begin{figure}[htb]
\centering
\includegraphics[width=3.3in,height=2.5in]{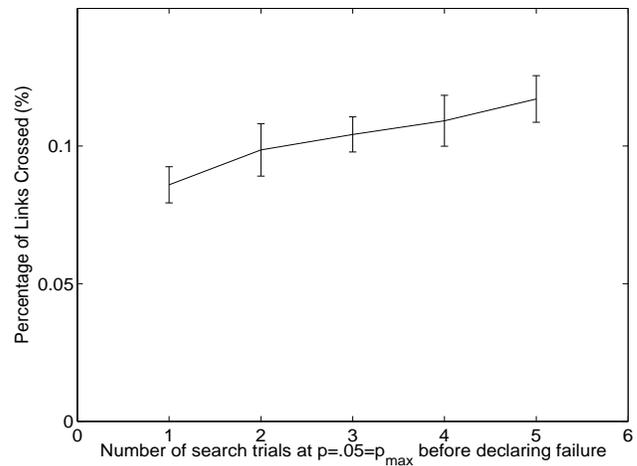}
\caption{\small {\bf Overall Traffic Per Query:} This figure plots
the percentage of edges crossed per query as a function of
 $n_{reps}$, average over 30 runs. Note that traffic cost is
extremely low (only around 0.1\% of network links need to be
crossed per query).  The error bar plots one standard deviation
plus and minus the mean.} \label{fig:frac_edges}
\end{figure}

Fig. \ref{fig:frac_edges}, plots the percentage of edges crossed
for per query as a function of $n_{reps}$, averaged over 30 runs.
(A query is defined as a series of percolation search trials as
defined in the subsection above.)  Note that the network traffic
cost is extremely low: on average, only approximately 0.1\% of the
84,190 email links in the network needs to be crossed in order to
get enough query hits to identify a suspected message as spam.
Combining results from Fig. \ref{fig:det_rate_plot} and Fig.
\ref{fig:frac_edges}, one can argue that $n_{reps} = 3$ is a good
operation point, since it gives near optimal spam detection
performance while incurring minimal traffic cost.

\begin{figure}[htb]
\centering
\includegraphics[width=3.3in,height=2.5in]{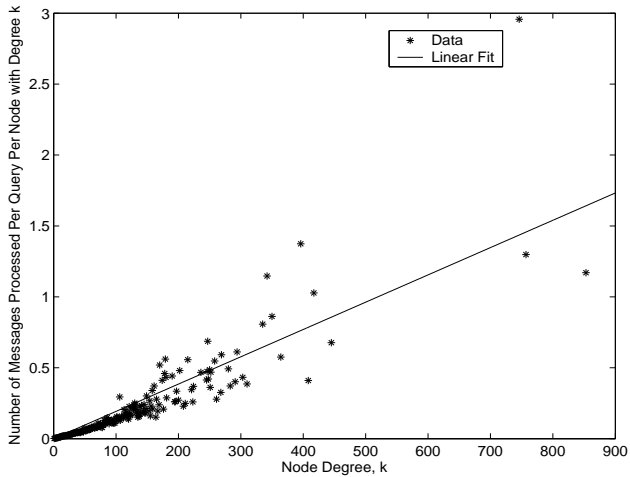}
\caption{\small {\bf Traffic vs. Degree:} The data points show the
average number of messages processed per percolation query for a
node with degree k (i.e. it is the total number of messages
processed per query for all nodes in the network with degree k
divided by the number of nodes with degree k.)  This plot is
obtained by using an $n_{reps}$ value of 3 for every percolation
query.  The slope of the linear fit is 0.0019 query/degree. since
each node forwards a query to a link with a fixed percolation
probability, we naturally expect that high-degree nodes handle
more messages.} \label{fig:traffic_plot}
\end{figure}

Fig. \ref{fig:traffic_plot} shows the network traffic as the
average number of messages processed by nodes with degree k.

\begin{figure}[htb]
\centering
\includegraphics[width=3.3in,height=2.5in]{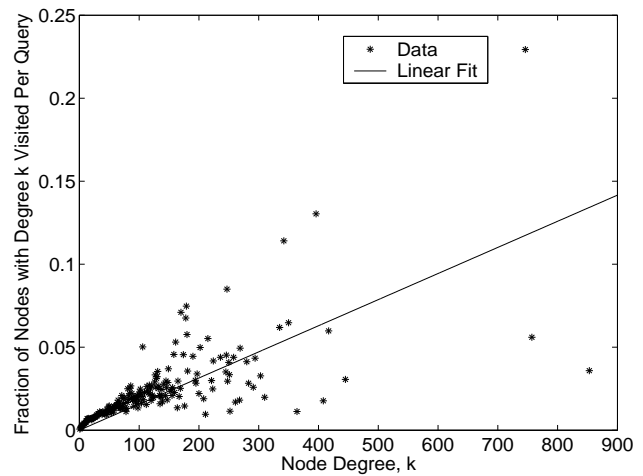}
\caption{\small The data points show the fraction of nodes with
degree k visited per percolation query (i.e. it is the number of
nodes with degree k visited per query divided by the total number
of nodes in the network with degree k.)  This plot is obtained by
using an $n_{reps}$ value of 3 for every percolation query.  The
slope of the linear fit is 1.573x$10^{-4}$/degree.}
\label{fig:frac_nodes}
\end{figure}

Fig. \ref{fig:frac_nodes} shows the average number of
participating nodes in a query as function of node degree. As
expected that high-degree nodes are more likely to be visited for
any given query since they are connected to a large number of
nodes.

\textbf{Bandwidth Cost Estimates:}  Fig. \ref{fig:frac_edges}
shows that the required traffic for each query is about 0.1\% of
edges, which corresponds to about 84 emails. Moreover, every short
email containing the digest of a message is about 1 KByte in size,
and every email incurs bandwidth cost on both the sender and
receiver. Thus, the bandwidth cost per query is approximately
(84+50) email exchanges (where the number 50 corresponds to the
random-walk query implantation with TTL=50), which at 1
KByte/email results in a total of 268 KByte per query. This total
traffic per query is distributed among all the nodes, and in
particular more among the high-degree nodes, as shown in Fig.
\ref{fig:traffic_plot}.

{\em Let us consider the worst case scenario first.} In the
network used for this simulation, a very high-degree node
typically processes around 1.5 messages per query in the network,
as seen from Fig. \ref{fig:traffic_plot}; only one set of nodes
uses more than this value.  Assuming that every user gets 1 spam
per hour, we conclude that a very high-degree node would need to
process about 85,500 messages per hour since there are around
57,000 nodes in the network.  Since the query message size is 1
KByte, the bandwidth cost on high degree nodes would be around 85
MByte per hour\footnote{In this calculation, the bandwidth cost
due to content caching and the routing back of messages is
ignored.}, which is equivalent to around 0.18 Mb/second.  For a
typical fast internet connection of 100 Mb/second, {\em this
represents about .2\% of bandwidth cost.}

For nodes with lower degree, {\em the cost is substantially
lower}.  For example, even a node with degree 100 would process on
the average 0.19 messages per query, and hence, using the same
estimate of 1 spam per node per hour, the bandwidth costs would be
only around 23K/second.

\section{Threat Model and Effectiveness of Trust Scheme}\label{sec:threat}

In this section, we will construct a model of malicious users in
the network trying their best to subvert the system. Through a
large-scale simulation, we will demonstrate that implementing the
EigenTrust algorithm presented in section \ref{ssec:trust} can
effectively reduce the damage inflicted by the malicious users.

With the system introduced so far, {\em a malicious node can
subvert the system by introducing blacklists of well-known valid
messages} into the network.\footnote{There are other possible
forms of attack such as a Denial-of-Service attack on the
high-degree nodes by artificially flooding the network with
queries. These forms of attacks are outside the scope of this
paper.} As a result, messages from mailing lists become easy
targets of an attacker. Note that this form of attack will only
raise the false positive rate
of the system and it has no impact on the spam detection rate.
%
Every malicious node will pick a fixed set of mailing lists and
periodically update the blacklist with new messages from the
mailing lists. In addition, it is assumed that the popularity of
mailing lists follows a Zipf distribution and the probability that
a mailing list is being queried follows the same Zipf law. We further
assume that the spammer wants to inflict maximum damage and thus
will select a given mailing list to blacklist following the same
Zipf distribution for popularity since users of the system are
more likely to be subscribed to popular mailing lists.

\subsection{Simulation Setup and Trust Scheme}
The simulation setup and parameters in this section will be
identical to the simulation performed in section
\ref{sec:simulation}, except for the following: {\em first}, a
small fraction of nodes in the network (250 nodes) will be
labelled as malicious nodes and these malicious nodes will
blacklist non-spams from popular mailing lists; {\em second}, for
simulation purpose, we assume that the probability that a node in
the email network is malicious is inversely proportional to its
in-degree, since low in-degree nodes are trusted by a few peer
email users and thus more likely to be malicious; {\em third}, the
malicious nodes will follow all specifications of the protocol
such as forwarding and routing queries, storing the cache implants
for other nodes, etc.\footnote{Malicious nodes can undermine the
performance of the system by failing to follow the protocol.
However, this is outside the scope of the paper.};  {\em fourth},
we relax the uniform spam arrival assumption in section
\ref{sec:simulation}.  In this simulation, the probability that a
node receives a spam is directly proportional to its in-degree.
The justification for this assumption is that a high in-degree
node signifies very active and long-time usage of the email
account and thus more likely to receive spams.  All relevant
parameters are specified in Table \ref{tab:sim_parameters}.

We then perform a Monte Carlo simulation on email network as
follows: at every time step, ten malicious nodes would insert
their malicious content, which consist of blacklists of non-spams;
also, 500 copies of the same spam message arrive as in section
\ref{sec:simulation}; In addition to the spam arrivals, a constant
number of non-spams would arrive and queried by users; based on
the hits that are routed back, nodes would classify the messages
queried to be spam or non-spam.

\begin{figure}[htb]
\centering
\includegraphics[width=3.3in,height=2.5in]{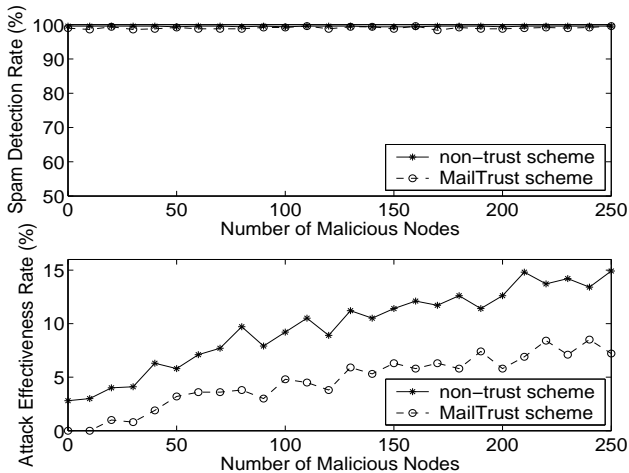}
\caption{\small {\bf MailTrust Performance:} The top and bottom
figures plot the spam detection rate and the attack effectiveness
rate as a function of the number of malicious nodes that have
joined the system. While both schemes yield approximately the same
detection rate, the MailTrust scheme results in a significantly
lower attack effectiveness rate. Note that we are assuming it is
easy for malicious nodes to know of messages (e.g., sent to
mailing lists) that a large number of nodes will receive. Clearly,
incorporating a white-list based scheme for processing messages
from mailing lists at the level of a {\em cyber}alter ego, would
be the best way of handling such attacks.} \label{fig:attack_sim}
\end{figure}

We will use two methods for spam classification: the {\em
non-trust scheme} and the {\em MailTrust scheme} (for
specification of the MailTrust algorithm, see section
\ref{ssec:trust}).  Under the non-trust scheme, a suspected
message is classified as spam if the number of distinct hits
routed back is greater than or equal to a threshold (the threshold
is set at 2 to give comparable performance as in section
\ref{sec:simulation}).  For the MailTrust scheme, the queried
message is identified to be spam if the sum of the MailTrust
scores of the distinct hits routed back is above a threshold.
This threshold is set to generate comparable spam detection rate
as the non-trust scheme. The results are plotted in Fig.
\ref{fig:attack_sim} for spam detection rate and false positive
rate as a function of the number of malicious nodes inserted.  As
shown in the plot, the malicious nodes have no impact on the spam
detection rate, since their blacklists of non-spams do not affect
the ability of other normal users to blacklist and identify spams.
From the spam detection rate plot, one can see that both schemes
generate comparable spam detection rates. However, by examining
the false positive rate plot, one can immediately see that {\em
the MailTrust scheme results in about 50\% improvement in lowering
the false positive rate.}

The reason for this improvement is mainly due to the fact that
high in-degree nodes tend to have high trust scores and receive
more spams. Thus, for the MailTrust scheme, we can set the
threshold value a little high and still have a very good spam
detection rate, because a large fraction of query hits for spams
will be provided by the high in-degree nodes who have high trust
scores.  In addition, most malicious nodes have low trust scores
since they tend to have low in-degrees (this assumption is made in
the subsection above).


\section{Miscellaneous}\label{sec:misc}

\subsection{Protection of Privacy}
\begin{figure}[htb]
\centering
\includegraphics[width=3.3in,height=2.5in]{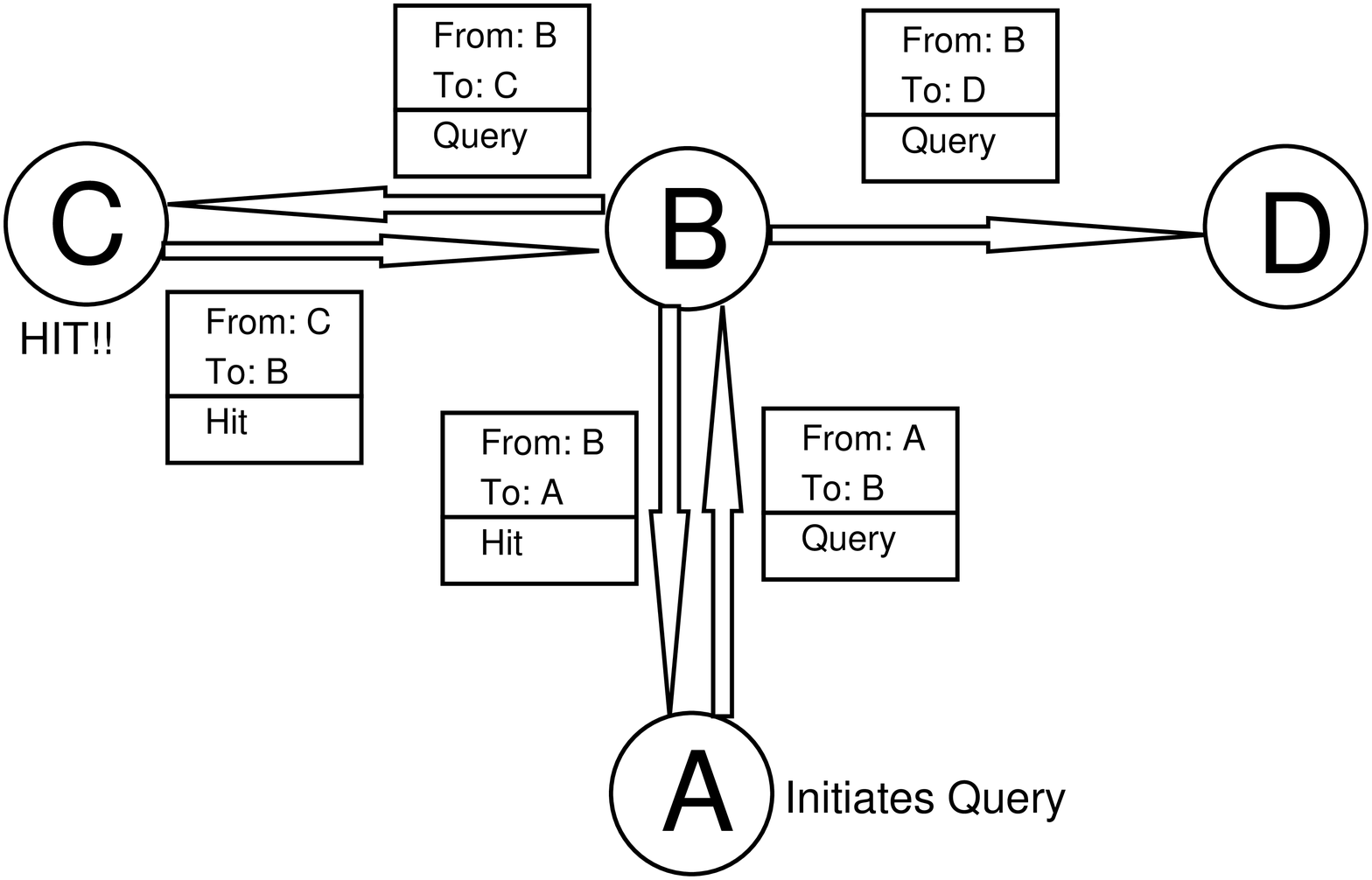}
\caption{\small {\bf Protecting Privacy:} A simple diagram
illustrating the secure version of message forwarding in the
system. For example, note that both node C and node D do not know
that the query comes from node A; similarly, node A does not know
that the hit comes from node C.} \label{fig:email_forward}
\end{figure}

Since our proposed anti-spam system is social network based, it is
very important to protect users' privacy by preventing anybody
from using the network to map out social links.  Furthermore, if a
malicious individual is able to map out the social email network,
a database of social contacts can be constructed to send out more
spams from spoofed personal contacts.  To address this problem,
all messages exchanged in the system must be forwarded
anonymously.  The basic idea is that when a node forwards a
message, any information pertaining to which nodes that the
message has visited must be deleted before forwarding. This keeps
all system communications to an acquaintance-acquaintance level,
Fig. \ref{fig:email_forward}.


%

\subsection{System Resilience against User Unreliability}
The users of the system are dynamic.  Namely, users will logon and
logoff as they wish.  Since our system heavily relies on the
underlying social email network, the natural question will be: how
many users in the system can be offline before the network is
severely segmented into many small components?

Alternately, we can re-phrase the above problem as followed: how
many nodes in a network can randomly fail before the network
becomes fragmented?  It turns out that this problem has been
extensively studied analytically and numerically \cite{Cohen,
Albert}. Using site percolation theory, Cohen et.al. \cite{Cohen}
shows that scale-free PL networks are extremely robust to random
failures: for a PL network with PL exponent less than 3, the
critical fraction of nodes, $p_c$, that needs to be removed for
the network to fragment goes to 1 as the network size approaches
infinity.  Furthermore, for a finite-size network with a large
number of nodes on the order of tens of thousands, the critical
fraction $p_c$ is well over 0.99. Since the social email network
is a PL network with exponent close to 2, these results from site
percolation theory is directly applicable.

Therefore, the network will not be fragmented even if a massive
number of system users suddenly leave. Alternately, one only needs
a very small fraction of the users to be using the system before
they can start successfully exchanging information.


\subsection{Simple Measures for Performance Improvement}

\textbf{Spam Traps.}  When our proposed system is deployed in the
real world, the initial number of users will be small.
In fact, all collaborative spam filters must overcome this "initial
hurdle" in order to become widely-used.

Our proposed solution to this "initial hurdle" problem is to
install \emph{spam traps}.  By definition, a spam trap is an email
account created for the sole purpose of attracting spams.  These
spam trap addresses can be easily promoted throughout the internet
to attract a large number of spams.  It has been noted by a
commercial anti-spam company that only a few hundred well-spread
spam traps are needed to catch almost all new spams
\footnote{http://www.esafe.com/pdf/esafe/\\esafe\_antispam\_whitepaper.pdf}.
These spam traps are not difficult to initiate and they do not
cost much in bandwidth and memory storage to maintain.  With spam
traps properly installed, the system is ready to be deployed and
offer superior spam detection performance.

\textbf{Hybrid and Multi Tier Design}. As discussed in section
\ref{sec:threat}, legitimate emails from popular mailing lists can
easily become blacklist targets of the malicious users of the
system.  In addition to the trust scheme we proposed in section
\ref{ssec:trust}, any traditional spam filtering technique can be
utilized as \emph{DefinitelySpam} and \emph{DefinitelyNotSpam}
function in Algorithm \ref{Alg:ProcessMail} to achieve enhanced
performance, and plug security holes in the collaborative system.

\section{Concluding Remarks}\label{sec:conc}
Our fairly comprehensive simulation results show that global
social email networks possess several properties that can be
exploited using {\em recent advances in complex networks theory}
(e.g., the percolation search algorithm) to provide an efficient
collaborative spam filter. Clearly, the proof-of-concept system
discussed here can be vastly improved
  and augmented with schemes that have proven successful at various
  levels. Moreover, there is nothing special about searching for and caching spam digests, and one can use our pervasive message passing system for managing a general distributed information system. The primary requirement is to be able to provide enough benefits to the users so that they are motivated to cooperate, which is relatively easily accomplished when it comes to spam management. If users get used to the spam filtering system, then we envision that queries for other information will follow.
%

The study brings out several aspects of the burgeoning cyberspace
networks, and the increasingly powerful {\em Cyber}alter ego: (i)
They have some of the same characteristics as their real-life
counterparts, and hence, can be managed and explored using
well-studied schemes; (ii) In many P2P applications, we do not
need to explicitly define new links and form the network from
scratch, but existing cyberspace and social contacts can be
exploited as an efficient P2P infrastructure.  Such existing
networks combined with efficient tools, borrowed from the field of
complex networks, can achieve almost optimum performance.  This
work and similar recent concepts \cite{Boykin} constitute some of
the first steps toward the management and design of efficient and
naturally grown collaborative systems in the cyberspace.

\bibliographystyle{abbrv}
\bibliography{p2p_antispam_ref}

\end{document}